\documentstyle[aps,prl,epsf,preprint]{revtex}
\def\fun#1#2{\lower3.6pt\vbox{\baselineskip0pt\lineskip.9pt
        \ialign{$\mathsurround=0pt#1\hfill##\hfil$\crcr#2\crcr\sim\crcr}}}

\begin{document}

\title{\vskip-2.5truecm{\hfill \baselineskip 14pt {{
\small  CERN-TH/96-149\\
       \hfill LPTHE-ORSAY 96/40\\
        \hfill hep-ph/9606342}}\vskip .1truecm} 
\vskip 0.1truecm {\bf D-Term Inflation }}
\author{{P.Bin\'etruy}\thanks{binetruy@qcd.th.u-psud.fr}$^{(1)}$ and 
{G.Dvali}\thanks{
 dvali@mail.cern.ch}$^{(2)}$}
\address{$^{(1)}${\it LPTHE, Universit\'e Paris-Sud, B\^at. 211, \\
F-91405 Orsay Cedex, France}}
\address{$^{(2)}${\it Theory Division, CERN\\Geneva, Switzerland}}
\maketitle


\begin{abstract}
\baselineskip 12pt

\end{abstract}
We show that inflation which is dominated by the $D$-term density
avoids the `slow-roll' problem of  inflation in  supergravity.
Such an inflationary scenario can naturally emerge
in theories with non-anomalous or anomalous $U(1)$ gauge symmetry.
In the latter case the scale of inflation is fixed by
the Green--Schwarz mechanism
of anomaly cancellation. The crucial point is
that the (super)gravity-mediated
curvature of all the scalar fields (and, in particular, of the inflaton),
which in the standard $F$-dominated case is of the order of the
Hubble parameter, is absent in the $D$-term
inflation case. The curvature of
moduli and of all other flat directions during such an inflation
crucially depends on their gauge charges.
\thispagestyle{empty}

\newpage
\pagestyle{plain}
\setcounter{page}{1}
\def\beq{\begin{equation}}
\def\eeq{\end{equation}}
\def\beqa{\begin{eqnarray}}
\def\eeqa{\end{eqnarray}}
\def\tr{{\rm tr}}
\def\x{{\bf x}}
\def\p{{\bf p}}
\def\k{{\bf k}}
\def\z{{\bf z}}
\baselineskip 20pt
Most models of inflation \cite{inf}
assume a period in the early history during
which the Universe was dominated by a potential energy of a slowly
rolling scalar field -- the inflaton. For this to happen,
certain slow-roll
conditions must be satisfied. These conditions ensure that the curvature
of the inflaton potential is smaller than the instant value of the Hubble
constant ($H$). Inflation stops whenever this is not the case.

According to the modern view, 
particle physics is described below Planck scale ($M_P$) energies
by an effective $N = 1$ supergravity theory.
As  was pointed by many authors\cite{infx},\cite{infh1}
\cite{infh2}, it is very difficult to naturally
implement a `slow-roll' inflation in the  context of supergravity.
The problem
has to do with the fact that 
inflation, by definition, breaks global supersymmetry since it requires
a non-zero cosmological constant $V$ (false vacuum energy of the inflaton).
However, in  supergravity theories, the supersymmetry breaking gets
transmitted to all the fields by the universal messenger which is 
gravity, and the resulting soft masses of the scalar fields are typically
\cite{infh1},\cite{infh2}\cite{modh}
\begin{equation}
m_{soft}^2 \sim {V \over M_p^2} \sim H^2.
\end{equation}
An unnatural fine-tuning is required in order to avoid a soft mass of the 
same magnitude for the inflaton itself. This is usually considered as a
generic problem for  inflation in the context of supergravity.

 The aim of the present letter is to  show that the above difficulty only
exists as far as inflation proceeds in a $D$-flat direction, with 
the vacuum energy being
dominated by the $F$-term(s) of some of the superfield(s);
it is not
true in the  case when the inflation is dominated by non-zero
$D$-terms ( as we will see below, this does not necessarily require the
inflaton to be a gauge-charged field ). The crucial difference is that
the gravity-mediated
soft scalar masses for the latter case are much smaller than
$H$. 
This happens because in the universe with positive cosmological constant
supersymmetry is broken, but the soft scalar masses
depend whether non-zero vacuum energy comes from the
$F$-term or $D$-term. In the $F$-type breaking the scalars are
getting soft masses given by (1), whereas for the
$D$-type breaking, these masses depend on their gauge charges:
scalars charged under the corresponding gauge symmetries gain masses
$>> H$, whereas others can only get gauge-mediated masses from the
loops\cite{dv}. In particular, for the inflaton (which is assumed to
be a singlet under gauge symmetries in question) the
curvature is automatically small.

This observation can have a crucial impact also for
the cosmology of the flat directions and, in particular, for the
cosmological moduli problem \cite{moduli}.
We consider a class of inflationary scenarios where inflation is
dominated by non-vanishing D-terms and we
show that they naturally emerge in theories with non-anomalous
or anomalous gauge $U(1)$ symmetry which include a Fayet-Iliopoulos
$D$-term. Such  scenarios can be considered as natural supersymmetric
realizations of the `hybrid' inflation invented by Linde \cite{hybrid}.
Below we will refer to the inflation dominated by $D$-terms as
$D$-inflation.

Before constructing an explicit example, let us first discuss the
problems of inflation in supergravity and show why they can be
avoided in the case of $D$-inflation. The two slow
rolling conditions for the potential $V$ of the
inflaton field $X$ can be defined in the
following form\cite{inf}
\begin{equation}
\left | {MV' \over V} \right | << 1
\end{equation}
and
\begin{equation}
\left | V'' \right | << H^2
\end{equation}
where prime denotes the derivative with respect to $X$ and $M = 
M_p/\sqrt{8\pi}$.
Breakdown of one of these conditions signals the end of inflation
\footnote{Needless to say, these conditions are not applicable if the
inflationary state is a local minimum. This
may happen if, for instance, the thermal effects force the Higgs
field
(which at zero temperature has a large VEV and
an almost flat potential) to be temporarily trapped at the origin, but the
vacuum energy dominates the thermal energy \cite{bg}. Such an inflation,
unfortunately, has too few e-foldings (and a very small $H$) in order to
solve the flatness and horizon problems.}.
Now, in supergravity, the scalar potential has the following
form\cite{sugra}
\begin{equation}
V = {\rm e}^{{K \over M^2}}\left[ (K^{-1})_i^j F_iF^j - 3 {|W|^2 \over M^2}
\right] + {g^2 \over 2}{\rm Re}f_{AB}^{-1}D^AD^B
\end{equation}
where
\begin{equation}
F^i = W^i + K^i {W \over M^2}
\end{equation}
and
\begin{equation}
D^A = K^i (T^A)_i^j\phi_j + \xi^A.
\end{equation}
Here $K$, $W$, and $f$ are respectively the K\"ahler potential,
superpotential
and
gauge-kinetic function. Upper (lower) indexes ($i,j$) denote
derivative with respect to field $\phi_i~(\phi^{*i})$ and $T^A$ are
generators of the gauge group in the appropriate representation. The
$\xi^A$ are Fayet-Iliopoulos $D$-terms, which can only exist for the
$U(1)$ gauge groups. In what follows the crucial role is played
by the exponential factor ${\rm e}^{{K \over M^2}}$
in front of the $F$-terms. The necessary condition for the `slow-roll'
inflation is the existence of a positive (false) vacuum energy,
meaning that at least some $\left < F_i \right >$ or some
$\left < D^A \right >$ are nonzero. Thus, inflation necessarily breaks
supersymmetry. Let us assume for a moment that the inflation is
dominated by some of the $F_i$ terms and that the $D^A$-terms are
vanishing or negligible (most of the existing  scenarios in the literature
assume this condition). Then the slow-roll conditions (2) and (3)
can be written as
\begin{equation}
{MV' \over V} = {K_X \over M} + {\rm other~terms} << 1
\end{equation}
and
\begin{equation}
V'' = 3 K^X_X H^2 + {\rm other~terms} << H^2
\end{equation}
respectively. Here the subscript $X$
denotes a derivative with respect to  the
inflaton  and  we have used
the relation $H^2 = V/3M^2$ to obtain the second equation. 
The `other terms' in these expressions
are typically of the same order as the ones written explicitly.
Their precise value is  model dependent, and they only cancel
in some special cases (e.g. if $X$ is a Goldstone
field\cite{gold}
or the dilaton or one of the moduli in some string models). 
Generically, however, this cancellation
requires a fine tuning, which we will
ignore here. In any case, neglecting
these `other terms'  gives the correct order of magnitude.

Now,  during inflation, unless a very special form is chosen,
$K_X$ is typically of the order of $X$. Thus, in principle, one can satisfy
equations (1) and (7) if inflation occurs for values well below the Planck
scale. This is difficult to achieve for the chaotic inflationary scenario
\cite{chaotic}. The second condition is even  more severe \cite{infh1},
\cite{infh2}.
The quantity $K_X^X$ stands in
front of the kinetic term and therefore in the
true vacuum it should be normalized to one. Then it is very unlikely
to expect it to be much smaller during inflation.
These arguments indicate that it is not easy to implement $F$-type
inflation in supergravity theories. All the
solutions proposed that we know of involve
specific non-minimal forms of the K\"ahler potential.
While it is not excluded that such forms indeed may emerge
from the underlying fundamental theory, in the present paper we will
study an alternative solution, which is largely independent on the
possible forms of the K\"ahler potential and   works equally well
for its minimal form ($K_X^X = 1$).

What is interesting about 
$D$-inflation is that the problems discussed above 
can be automatically avoided.
Indeed for  inflation dominated by some of the $D^A$-terms the
slow-roll conditions can be easily satisfied: quantities $(D^A)_X$ and
$(D^A)_X^X$ can be automatically zero and what one needs, just, is
that $X$ is annihilated by the corresponding gauge generator.
 Some other important differences emerge: first, the
slope
of the inflaton potential is induced from the one-loop radiative
corrections and is practically independent of the details of
supersymmetry breaking in the present vacuum;
secondly, the inflation-induced curvature of the moduli
and other flat directions crucially depends on their gauge-charges.

Let us show that such a scenario can naturally emerge in a theory with
a $U(1)$ gauge symmetry. First we consider an example with a non-anomalous
$U(1)$ symmetry. We introduce three chiral superfields $X$, $\phi_+$ and
$\phi_-$ with charges equal to $0$, $+ 1$ and $- 1$ respectively.
The superpotential has the form
\begin{equation}
W = \lambda X\phi_+\phi_-
\end{equation}
which can be justified by several choices of discrete or continuous
symmetries and in particular by $R$-symmetry.
The scalar potential in the global supersymmetry limit reads:
\begin{equation}
V = \lambda^2 |X|^2 \left(|\phi_-|^2 + |\phi_+|^2 \right) +
\lambda^2|\phi_+\phi_-|^2 + 
{g^2 \over 2} \left(|\phi_+|^2 - |\phi_-|^2  + \xi \right)^2
\end{equation}
where $g$ is the gauge coupling and $\xi$ is a Fayet-Iliopoulos $D$-term
(which we choose to be positive).
This system has a unique supersymmetric vacuum with broken
gauge symmetry
\begin{equation}
X = \phi_+ = 0, ~~~ |\phi_-| = \sqrt{\xi}.
\end{equation}
Minimizing the potential, for  fixed values of $X$, with respect to
other fields, we find that for  $|X| > X_c = {g \over \lambda}
\sqrt{\xi}$, the minimum is at $\phi_+ =\phi_- = 0$. Thus, for
$|X| > X_c$ and $\phi_+ =\phi_- = 0$ the tree level potential
has a vanishing curvature in the $X$ direction and large positive
curvature in the remaining two directions:
\begin{equation}
m_{\pm}^2 = \lambda^2|X|^2 \pm g^2\xi.
\end{equation}
For arbitrarily large $X$ the tree level value of the potential remains
constant: $V = {g^2 \over 2}\xi^2$. Thus $X$ is a natural inflaton.
Indeed, under the assumption of  chaotic initial conditions
$|X| >> X_C$ this system naturally
leads to  inflation. Along the inflationary
trajectory all the $F$-terms vanish and Universe is dominated by the
$D$-term which splits the masses of the Fermi-Bose components in the
$\phi_+$ and $\phi_-$ superfields. Such splitting results in a
one-loop effective potential for the inflaton field\cite{dss},
In the present case this potential can be easily evaluated and for large
$X$ it behaves as
\begin{equation}
V_{eff} = {g^2 \over 2}\xi^2 \left( 1 + {g^2 \over 16\pi^2} {\rm ln}
{\lambda^2 |X|^2 \over \Lambda^2}\right)
\end{equation}
Slow-roll conditions break down and inflation stops
when $X \sim \sqrt{C}M$, where $C$ is a one loop factor of order
$g^2/(16 \pi^2)$.
In the globally supersymmetric limit, the dynamics of the above
scenario is somewhat similar to the `hybrid scenario'' of
\cite{dss}, but it exhibits a crucial
difference in the locally supersymmetric
case: due to its $D$-type nature it escapes the problems of the standard
$F$-type inflation. To demonstrate this explicitly let us consider the
supergravity extension of our model. For definiteness we will assume 
the minimal structure for $f$ and the K\"ahler potential
\begin{equation}
K = |\phi_-|^2 + |\phi_+|^2 + |X|^2.
\end{equation}
Note that this form maximizes  the problems for the $F$-type
inflation since it gives $K_X^X = 1$. The scalar potential reads
\begin{eqnarray}
V &=& {\rm e}^{|\phi_-|^2 + |\phi_+|^2 + |X|^2 \over M^2}  
\lambda^2 \left[|\phi_+\phi_-|^2 \left(1 + {|X|^4 \over M^4} \right) 
\right. \nonumber \\
& & \; \; \; \; \; \; \; \; \; 
\left. + |\phi_+X|^2 \left(1 + {|\phi_-|^4 \over M^4} \right)
+ |\phi_-X|^2 \left(1 + {|\phi_+|^4 \over M^4} \right)
+ 3{|\phi_+\phi_-X|^2 \over M^2}\right] \nonumber \\
&+& {g^2 \over 2} \left(|\phi_+|^2 - |\phi_-|^2  + \xi \right)^2 
\end{eqnarray}
Again for  values of $|X| > X_C$,  other fields than $X$ vanish and 
the  behaviour is much similar to   the global supersymmetry case.
The zero tree level curvature of the inflaton potential is not affected
by the exponential factor in front of
the first term in (15), since this term is vanishing
during inflation. This solves the slow-rolling problems of the $F$-type
inflation.

 Let us now consider the case of an
anomalous $U(1)$ symmetry. Such symmetries
usually appear in the context of string theories \cite{dsw} and the
anomaly cancellation is due to the Green--Schwarz (GS)
mechanism\cite{gs}, which determines the
value of the Fayet-Iliopoulos D-term as
\begin{equation}
\xi_{GS} = {{\rm Tr}Q \over 192\pi^2} g^2 M^2
\end{equation}
where the trace is taken over all charges.
To see how  $D$-inflation is realized in this case,
let us consider the simple example of such a
$U(1)$ symmetry under which
$n_+$ chiral superfields $\phi_+^i$ and $n_-$ superfields
$\phi_-^{A}$ carry one unit of positive and negative charges respectively.
For definiteness let us assume that $n_+ > n_-$, so that the symmetry is
anomalous and $Tr{Q}\neq 0$.
We assume that some of the fields transform under
other gauge symmetries, since the GS mechanism requires non-zero mixed
anomalies. Let us introduce a single gauge-singlet superfield $X$.
Then the most general trilinear coupling of $X$ with the charged
superfields can be put in the form:
\begin{equation}
W = \lambda_A X\phi_+^A\phi_-^A
\end{equation}
(for simplicity we assume additional symmetries that forbid direct mass
terms). Thus there are $n_+ - n_-$ superfields with positive charge
that are left out of the superpotential. The potential has the form
\begin{equation}
V = \lambda_A^2 |X|^2 \left(|\phi_-^A|^2 + |\phi_+^A|^2 \right) +
\lambda_A^2|\phi_+^A\phi_-^A|^2 + 
{g^2 \over 2} \left(|\phi_+^i|^2 + |\phi_+^A|^2 - 
|\phi_-^A|^2  + \xi_{GS} \right)^2
\end{equation}
where summation over $A= 1,2...,n_-$ and $i = 1,2...,(n_+ - n_-)$ is
assumed. Again, minimizing this potential for  fixed values of
$X$ we find that for $|X| > X_C = max\left({g \over \lambda_A}\right)
\sqrt{\xi_{GS}}$, the minimum for all $\phi_+$ and $\phi_-$ fields is
at zero. Thus, the tree level curvature in the $X$ direction is
zero and inflation can occur. During inflation masses
of $2n_-$ scalars are
\begin{equation}
m_{\pm}^2 = \lambda_A^2|X|^2 \pm g^2\xi_{GS}.
\end{equation}
and the remaining $n_+ - n_-$ positively charged scalars have masses squared
equal to $g^2\xi_{GS}$. We see that  inflation proceeds much in the same
way as for the non-anomalous $U(1)$ example discussed above. The interesting
difference is that in the latter case the scale of inflation is an arbitrary
input parameter (although in concrete cases it can be determined by the
GUT scale), whereas in the anomalous case it is predicted by
the GS mechanism.

Generalization of the above example to the case of fields with 
different charges and with several singlet
field is straightforward. The
inflaton field in such a case may not be a single gauge-singlet, but
rather a combination of several fields. Such trajectories are in general
model-dependent and must be studied case by case.
 A generic rule which leads to our scenario is that there must exist a 
$U(1)$-neutral direction in field space which does not appear alone in
the F-terms and along which all states
having charges opposite to $Tr{Q}$ gain
positive masses.

We think that the proposed mechanism may provide a somewhat generic
inflation scenario in the context of superstring models. In such models,
the presence of an anomalous $U(1)$ symmetry is not unusual; in general,
several fields are neutral under the $U(1)$ symmetry and thus they (or a
combination of them) may play the role of $X$. The presence of several
other non-anomalous $U(1)$ symmetries may prove to be useful in order to
prevent $X$ from contributing alone to the F-terms (say a term $\lambda
X^3$ in our simple model presented above). The requirement of
cancellation of mixed anomalies through the Green-Schwarz mechanism may
provide stringent constraints on the charges of such $U(1)$, constraints
which are a priori automatically accounted for in string models.

 Finally let us note some other crucial differences between the $D$-type
and $F$-type inflations. Flat  directions behave very differently 
during inflation.
In the ordinary $F$-inflation case all flat directions, including moduli,
are getting curvature at least of order $H$ from the gravity mediated
supersymmetry breaking (note that for the gauge-charged flat directions
curvature can be much larger due to gauge-mediated breaking\cite{dv}).
This mass is exactly of the same origin as the problematic
mass for the inflaton and appears due to the exponential factor
${\rm e}^{K/M^2}$ in front of the $F$-terms. Obviously,
no such contribution exists in the $D$-inflation scenario outlined above.
So in this case the curvature of the flat directions during inflation
crucially depends on their gauge charge. In this respect we can divide
all flat directions in three categories:
1) The flat directions which carry no common charge with any of the fields
charged under $U(1)$. Such flat directions will not receive any
contribution to the curvature, at least up to a three loop order.
2) Flat
directions that are neutral
under $U(1)$, but carry some gauge charges
also carried by the $U(1)$-charged fields.
Such direction will receive
a two-loop gauge-mediated soft masses \cite{dv};
for example, if some $\phi_+^A$ and $\phi_-^A$ superfields are transforming
as fundamental and anti-fundamental representations of some $SU(N)$
gauge group, then all $SU(N)$ nonsinglet fields during inflation
will get the universal
(up to charges) two-loop gauge-mediated soft masses
\begin{equation}
m_{soft}^2 \sim \left( \alpha \over 4\pi \right)^2{g^4 \over \lambda_A^2}
{\xi^2 \over |X|^2} \sim \left( \alpha \over
4\pi \right)^2{g^2 \over \lambda_A^2} {M^2 \over |X|^2} H^2
\end{equation}
where $\alpha$ is a gauge coupling of $SU(N)$.
3) Directions charged under $U(1)$ receive large tree level masses.
 
We expect that all this will have
important consequences for the cosmological
moduli problem and the Affleck-Dine scenario\cite{ad} of baryogenesis.

\begin{enumerate}

\end{enumerate}

\end{document}